\documentstyle[preprint,revtex,12pt]{aps}
\begin{document}
\draft
\noindent
\begin{centering}

{\Large	FIRST-ORDER LAGRANGIANS AND PATH-INTEGRAL QUANTIZATION 
IN THE t-J MODEL.\\}

\vspace{1.cm}
{\bf A. Foussats, A. Greco and O. S. Zandron}\\

\vspace{1cm}

{\em Facultad de Ciencias Exactas Ingenier\'{\i}a y Agrimensura e IFIR 
(UNR-CONICET)\\
Av.Pellegrini 250 - 2000 Rosario - Argentina.}\\

\end{centering}

\begin{abstract}

\begin{centering}
{\bf Abstract\\}
\end{centering}

By using the supersymmetric version of the Faddeev-Jackiw symplectic 
formalism, a family of first-order
constrained Lagrangians for the t-J model is found. In this approach the
Hubbard ${\hat X}$-operators are used as field variables.
In this framework, we first study the spinless fermion model which
satisfies the graded algebra spl(1,1).
Later on, in order to satisfy the Hubbard ${\hat X}$-operators commutation
rules satisfiying the graded algebra spl(2,1), 
the number and kind of constraints that must
be included in a classical first-order Lagrangian formalism for the t-J
model are found. This model is also analyzed in the context of the path- 
integral formalism, and so the correlation generating functional and the 
effective Lagrangian are constructed.

\end{abstract}


\narrowtext
 
\section{Introduction}

In a recent paper\cite{1} the classical and quantum Lagrangian dynamics
for the SU(2) bosonic algebra was constructed. By means of the
path-intergral techniques, the perturbative formalism
was also developed. In that case the
Hubbard operators\cite{2} are boson-like one, and they are a suitable 
representation for the spin-$1/2$ Heisenberg model.

The well known t-J model is one of the most important candidate to explain
the phenomenology of High-Tc superconductivity. This model contains the
main physics of doped holes on an antiferromagnetic background.
In the case of the t-J model, the Hubbard operator representation is quite
natural to treat the electronic correlation effects\cite{3}. In this model
in which spin and charge degrees of freedom are present, the Hubbard 
${\hat X}$-operators satisfies the graded algebra spl(2,1)\cite{4}.

Like as it was shown in Ref.[1] for the bosonic case, it must be expected that
the path-integral formalism applied to the t-J model described in terms of
a first-order Lagrangian can be useful. As it is well known, these techniques 
are powerful in quantum field theory as well as in solid state physics.  
This is clearly proved when the perturbative formalism  can be assumed, 
and consequently the Feynman rules and the diagrammatics of the model can be 
implemented.

In the t-J model the only three possible states on a lattice site are 
$\mid\alpha>$ = $\mid 0>$, $\mid +>$, $\mid ->$. These states correspond
respectively to an empty site, an occupied site with an electron of spin-up, 
or an occupied site with an electron of spin-down. Double occupancy is 
forbiden in the t-J model. In terms of these states the Hubbard 
${\hat X}$-operators are defined as\cite{2},

\begin{equation}
{\hat{X}}^{\alpha \beta}_{i} = \mid i \alpha>< i \beta\mid\; .
\eqnum{1.1}  
\end{equation}

In Eq.(1.1), when one of the index is zero and the other different 
from zero, the corresponding ${\hat X}$-operator is fermion-like, otherwise 
boson-like. 

The Hubbard ${\hat X}$-operators satisfy the following graded commutation 
relations

\begin{equation}
[{\hat{X}}_{i}^{\alpha \beta}\,,\,{\hat{X}}_{j}^{\gamma \delta}]_{\pm}
=\delta_{ij}(\delta^{\beta \gamma}{\hat{X}}_{i}^{\alpha \delta}\; {\pm}\;
\delta^{\alpha \delta}{\hat{X}}_{i}^{\gamma \beta})\; , 
\eqnum{1.2}
\end{equation}

\noindent
where the $+$ sign must be used when both operators are
fermion-like, otherwise it corresponds the $-$ sign.

Using the ${\hat X}$-operators definition it is easy to see that the following
conditions hold:

\noindent
a) the completeness condition

\begin{equation}
\sum_{\alpha}{\hat{X}}_{i}^{\alpha \alpha} = {\hat I}\;,  
\eqnum{1.3}
\end{equation}
 
\noindent
b) the multiplication rules in a given site

\begin{equation}
{\hat{X}}_{i}^{\alpha \beta}{\hat{X}}_{i}^{\gamma \delta} = 
\delta^{\beta \gamma}{\hat{X}}_{i}^{\alpha \delta}\; . 
\eqnum{1.4}
\end{equation}

One of the purpose of this paper is to construct a family of first-order 
Lagrangians
written in terms of fermion-like and boson-like Hubbard ${\hat X}$-operators, 
describing the dynamics of the t-J model. 

The graded commutators between the field dynamical 
variables i.e., the graded quantum Dirac brackets\cite{5} of the model must 
verify the
graded algebra spl(2,1) given in (1.2) for the Hubbard ${\hat X}$-operators.

This problem will be treated by extending our results of Ref.[1] to the
t-J model case. To this aim we will use the supersymmetric
extension\cite{6,7,8} of the symplectic Faddeev-Jackiw (FJ) method\cite{9}.   

Subsequently, once the family of Lagrangians and the constraint structure 
of the model were determined, by using the path-integral formalism the 
correlation generating functional can be found in terms of a suitable 
effective Lagrangian.   

The paper is organized as follows. In section II, the constraint structure
of the spinless fermion case is 
briefly analyzed in the framework of the symplectic FJ Lagrangian method. 
Next, by using the path-integral representation, the partition 
function is
written in terms of an effective Lagrangian. The final expression we find 
agrees perfectly with the current form of the   
partition function for the spinless fermion model, showing the validity of 
our approach.  
In section III, a general treatment for systems containing the Hubbard 
${\hat X}$-operators of the graded algebra spl(2,1) as dynamical variables is 
constructed. A family of classical first-order Lagrangians describing these 
dynamical systems is found. In section IV, using these results and by applying 
techniques used in quantum field 
theories, the path-integral quantization formalism is developed and the
correlation generating functional is given in terms of an appropriate 
effective Lagrangian. Conclussions are written in section V.

\section{The spinless fermion case}

The treatment of the spinless fermion model is useful for two main reasons: 
a) to understand how the algorithm is applied when both, boson-like and 
fermion-like Hubbard ${\hat X}$-operators are present; and b) to show that 
our approach produces the correct and well known path-integral representation
for the correlation generating functional of the model. 

By generalizing the results of Ref.[1], our purpose is to construct a family
of classical first-order Lagrangians written in terms of the Hubbard 
${\hat X}$-operators whose graded commutation relations or graded quantum 
Dirac brackets between field variables are given by the Eq. (1.2). 
This approach clearly shows how the constraint structure i.e.,  
the kind and the number of constraints present in these models is 
provided by the symplectic FJ quantum method. 

The notation and key equation related to the symplectic FJ method we will 
use are those of Refs. [6,7].

We assume that the family of classical first-order Lagrangians in terms of 
the Hubbard ${\hat X}$-operators can be written as follows 

\begin{equation}
L = a_{\alpha \beta}(X){\dot{X}}^{\alpha \beta} - {\bf V^{(0)}}\;.
\eqnum{2.1}
\end{equation} 

In the FJ language the symplectic potential ${\bf V^{(0)}}$ is defined by

\begin{equation}
{\bf V^{(0)}} = H(X) + \lambda^{a}\Omega_{a}\;, 
\eqnum{2.2}
\end{equation}

\noindent
and so the constraints are given by

\begin{equation}
\frac{\partial{\bf V}^{(0)}}{\partial{\lambda^{a}}} = \Omega_{a}\;.
\eqnum{2.3}
\end{equation}

Looking at the Lagrangian Eq. (2.1) we see that the initial set of
dynamical symplectic variables defining the extended configuration space
is given by $(X^{\alpha \beta}\;,\;\lambda^{a})$. 

In Eq. (2.2), $H(X)$ is a proper Hamiltonian for the model 
also given in terms of the Hubbard $X$-operators. It is important to 
remark that at this level the $X$-variables must be treated as classical 
fields.

The site subscript indices $i,j,$ appearing in the definition of the
Hubbard operators were dropped since they are irrelevant in the analysis 
we will develop. Without any difficulty the site indices can be opportunely 
included. 

In Eq. (2.1) the coefficients $a_{\alpha \beta}(X)$ are a priori 
unknown and they are determined in such a way that the graded algebra (1.2) 
for the  Hubbard $X$-operators must be ve\-ri\-fied. The reality condition on 
the Lagrangian implies that $a_{\alpha \beta}(X)$ = $ (-1)^{\mid a \mid}
a^{*}_{\beta \alpha}(X)$, where ${\mid a \mid}$ indicate the Fermi grading of 
the coefficient. The $\lambda^{a}$ parameters used in Eq.(2.2) are suitable 
bosonic or fermionic Lagrange multipliers which allow the introduction of the 
constraints in the Lagrangian formalism. $\Omega_{a}(X)$ is the set of unknown 
bosonic or fermionic constraints, initially considered ad hoc in the 
Lagrangian. Both the constraints $\Omega_{a}(X)$ as well as the range of the 
index $a$ (i.e., kind and number of contraints) must be determined later on 
by consistency. 

By following the steps developed in Refs.[7,8] the FJ method must be
implemented on the Lagrangian (2.1), and so  
the symplectic supermatrix $M_{AB}(X)$ can be constructed straightforward.
The matrix elements of the symplectic supermatrix are 
given by the generalized curl constructed with the partial derivatives
involving the set of variables.

When the symplectic supermatrix $M_{AB}(X)$ is singular it is necesary
to carry out the iteration procedure by enlarging the configuration space. 
Otherwise, when the symplectic supermatrix $M_{AB}(X)$ 
is non-singular, the inverse supermatrix $(M^{AB})^{-1}$ is unique and their 
matrix elements are the generalized FJ brackets,  
corresponding to the graded Dirac brackets of the theory. 

As usual the relation between the graded Dirac Brackets $\{\;,\;\}$   
and the graded commutation relations (2.1) for the
Hubbard $X$-operators is: $i\{X\,,\,X\}_{\pm}\rightarrow[{\hat
X}\,,\,{\hat X}]_{\pm}$.
   
As it is well known
the main feature of the symplectic formalism is that the classification of
constrained or unconstrained systems is related to the singular or non
singular behavior of the fundamental symplectic two-form which gives rise
to the symplectic supermatrix.
On the contrary to the Dirac's language, the classification of constraints in
primary, secondary and so on; or in first-class and second-class has no
meaning. 

On the other hand, when the FJ symplectic method is applied to gauge
models having constraints associated to gauge symmetries,
the algorithm is unable to produce an invertible
symplectic supermatrix. Therefore, the existence of the inverse of the
symplectic supermatrix necessarily implies that the model has only constraints 
which in the Dirac picture correspond to second-class one.

Now we are going to apply the symplectic formalism to the 
spinless fermion model. As it is known this model is obtained from the graded 
algebra (1.2) when the 
indices $\alpha\,,\,\beta$ can take only two values: $0$ denoting an empty 
site, 
and $1$ denoting an occuped site with one fermion. So, the four Hubbard 
${\hat X}$-
operators close the graded algebra spl(1,1)\cite{10}. Two of them 
${\hat X}^{11}$ and ${\hat X}^{00}$ are boson-like operators, 
while the other two  ${\hat X}^{01}$ and ${\hat X}^{10}$  are 
fermion-like one.

It is easy to show that in this case the symplectic supermatrix 
obtained from the Lagrangian (2.1) in which the field variables are 
$(X^{\alpha \beta}\;,\;\lambda^{a})$ is singular, and so it is necessary 
to carry out 
one iteration to obtain an invertible symplectic supermatrix.
As commented above, this is done by enlarging the configuration space 
redefining the $\lambda^{a}$ variables as: $\lambda^{a} = - {\dot{\xi}^{a}}$.
  
Consequently, after the iteration is done the constraints are written 
in the symplectic part and the first-iterated Lagrangian $L^{(1)}$
writes as follows  

\begin{equation}
L^{(1)} = a_{11}(X){\dot{X}}^{11} + a_{00}(X){\dot{X}}^{00} +
a_{01}(X){\dot{X}}^{01} + a_{10}(X){\dot{X}}^{10} +  {\dot{\xi}^{a}}
\Omega_{a} - {\bf V^{(1)}}\;.  
\eqnum{2.4}
\end{equation}

\noindent
where ${\bf V^{(1)}} = {\bf V^{(0)}}\mid_{\Omega_{a} = 0}\; = H(X)$. 

Therefore, the modified symplectic supermatrix associated to the 
Lagrangian (2.4) can be formally written

\begin{eqnarray}
M_{AB} =
\left( \matrix {
\frac{\partial a_{\gamma \delta}}{\partial X^{\alpha \beta}}
- \frac{\partial a_{\alpha \beta}}{\partial X^{\gamma \delta}}
&\frac{\partial\Omega_{b}}{\partial X^{\alpha \beta}}   \cr
- \frac{\partial\Omega_{a}}{\partial X^{\gamma \delta}}    &0    \cr }
\right) \; , \eqnum{2.5}
\end{eqnarray}

\noindent
where the compound indices $A = \{{(\alpha \beta)}, a \}$ and 
$B =\{{(\gamma \delta)}, b\}$ run in the different ranges of the complete
set of variables defining the extended configuration space.

Now, in order to obtain an invertible symplectic supermatrix, the problem is 
to determine both, the Lagrangian coefficients  $a_{\alpha \beta}(X)$ and 
how many constraints ${\Omega_{a}}$ are provided by the symplectic FJ 
algorithm. Finally, by solving the equations on the constraints the functions 
${\Omega_{a}}(X) = 0$ must be found.
   
Accounting that the inverse of the symplectic supermatrix $M_{AB}(X)$ is 
written 

\begin{eqnarray}
(M^{AB})^{-1} =
\left( \matrix {
\{X^{\alpha \beta}\;,\;X^{\gamma \delta}\}   
&\{X^{\alpha \beta}\;,\;\xi^{b}\} \cr
\{\xi^{a}\;,\;X^{\gamma \delta}\}   &\{\xi^{a}\;,\;\xi^{b}\}    \cr }
\right) \; , \eqnum{2.6}
\end{eqnarray}

\noindent
each matrix element of the submatrix 
$\{X^{\alpha \beta}\;,\;X^{\gamma \delta}\}$
must be  equalled to each one of the
following Hubbard commutation relations of the spl(1,1) graded algebra 
 
\begin{eqnarray}
& \{X^{00}\;,\;X^{00}\}_{-} & = \{X^{00}\;,\;X^{11}\}_{-} = 
\{X^{11}\;,\;X^{11}\}_{-} = 0 \;, \nonumber \\ 
& \{X^{00}\;,\;X^{01}\}_{-} & = - i X^{01}\;,\;\{X^{00}\;,\;X^{10}\}_{-} 
= i X^{10}\;,
\nonumber \\
& \{X^{11}\;,\;X^{10}\}_{-} & =  - i X^{10}\;,\;
\{X^{11}\;,\;X^{01}\}_{-} = i X^{01}\;, \nonumber \\
& \{X^{01}\;,\;X^{10}\}_{+} & = - i (X^{00} + X^{11})\;.
\eqnum{2.7}
\end{eqnarray}
 
Of course in the Eq.(2.6) the three remaining submatrices
$\{X^{\alpha \beta}\;,\;\xi^{b}\}$, $\{\xi^{a}\;,\;X^{\gamma \delta}\}$ and
$\{\xi^{a}\;,\;\xi^{b}\}$ are unknown.

It is easy to show that the invertible symplectic
supermatrix $M_{AB}(X)$ given in Eq. (2.5) is a square $6\times6$
dimensional one, and it can be written in the form

\begin{eqnarray}
M_{AB} =
\left( \matrix {
A_{bb}     &B_{bf}   \cr
C_{fb}     &D_{ff}   \cr }
\right) \; , \eqnum{2.8}
\end{eqnarray}

\noindent
whose Bose-Bose parts $A_{bb}$ and Fermi-Fermi parts $D_{ff}$ are even 
elements of
the Grassmann algebra and whose Bose-Fermi parts $B_{bf}$ and Fermi-Bose parts
$C_{fb}$ are odd elements. As it is well known\cite{11} the inverse 
$(M^{AB})^{-1}$
exists if and only if $A_{bb}$ and $D_{ff}$ have an inverse.

In the present case the Bose-Bose part $A_{bb}$ is an ordinary
non-singular $4\times4$ dimensional 
matrix and the Fermi-Fermi part $D_{ff}$ is an ordinary non-singular 
$2\times2$ dimensional matrix.
  
By using the equation $M_{AB}(M^{BC})^{-1} = \delta^{C}_{A}$, and taking
into account the equation 

\begin{equation}
(M^{BC})^{-1} = -i (-1)^{\mid\varepsilon_{B}\mid}
\left[{\hat B}\;,\;{\hat C}\right]_{\pm}\;,
\eqnum{2.9}
\end{equation}

\noindent
where ${\mid\varepsilon_{B}\mid}$ is the
Fermi grading of the variable $B$, differential equations on 
the Lagrangian coefficients $a_{\alpha \beta}(X)$ and on 
the constraints $\Omega_{a}$ are obtained. 

In particular, the system of four homogeneous differential equations on the 
constraints ${\Omega_{a}}$ can be written
  
\begin{equation}
i (-1)^{{\mid \varepsilon_{\alpha \beta}\mid}({\mid a \mid} + 1)} 
\frac{\partial \Omega_{a}}{\partial X^{\alpha \beta}} 
\left[ X^{\alpha \beta}, X^{\gamma \delta}\right]_{\pm} =  0\; ,
\eqnum{2.10}
\end{equation}

\noindent
where ${\mid \varepsilon_{\alpha \beta}\mid}$ and ${\mid a \mid}$ are  
the Fermi grading of the Hubbard operators and of the constraints respectively.

By solving the partial differential equations system (2.10), two bosonic 
solutions are found. So, the associated constraints reads 

\begin{equation}
\Omega_{1} = X^{00} + X^{11} - 1 = 0 \;, 
\eqnum{2.11a}
\end{equation}

\begin{equation}
\Omega_{2} =  X^{11} + X^{01} X^{10}  - 1 = 0 \; .
\eqnum{2.11b}
\end{equation}

We emphasize that once the invertibility of the symplectic supermatrix (2.5)
is assumed, necessarily it must be understood that the constraints (2.11) are  
second-class one, as really occurs. 

Analogously, by solving the remaining system of partial differential 
equations on the Lagrangian coefficients $a_{\alpha \beta}(X)$, the values 
we find are

\begin{equation}
a_{00} = \frac{1}{2} X^{11}\;,\;a_{11} = - \frac{1}{2} X^{00}\;,\; 
a_{10} = \frac{i}{2} X^{01} \;,\; 
a_{01} = \frac{i}{2} X^{10}\;. 
\eqnum{2.12}
\end{equation}

The constraint (2.11a) provided by the symplectic FJ algorithm is precisely 
the completeness condition (1.3) which is necessary because 
the "double occupancy" at each site is forbiden.   
   
Therefore, from the Lagrangian (2.1) with the values of the coefficients given in 
(2.12), together with the bosonic constraints (2.11), 
it is straightforward to prove that the graded algebra spl(1,1) given in
Eq. (2.7) is recovered.

At this stage, we are able to write the correlation generating functional
by using the path-integral approach of Faddeev-Senjanovich\cite{12}. 
Thus, the spinless fermion model partition function can be initially written 
as follows

\begin{equation}
Z = \int {\cal D}X\; \delta(X^{01}X^{10} + X^{11} - 1)\;
\delta(X^{11} + X^{00} - 1) \; 
 exp\;i \int dt\; L(X,\dot{X}) \;,
\eqnum{2.13}
\end{equation}

\noindent
where the Lagrangian $L(X,{\dot X})$ is given by

\begin{equation}
L = \frac{1}{2}\left(X^{11}{\dot{X}}^{00} - X^{00}{\dot{X}}^{11}\right) +
\frac{i}{2} \left(X^{01}{\dot{X}}^{10} + X^{10}{\dot{X}}^{01}\right)
 - H(X)\;.
\eqnum{2.14}
\end{equation} 
 
We note that in the path-integral (2.13) the superdeterminant
of the supermatrix constructed from the constraints was omitted   
because it is field independent and can be included in the path-integral 
normalization factor.

By integrating out the bosonic variables $X^{11}$ and $X^{00}$, the
partition function (2.13) becomes

\begin{equation}
Z =  \int {\cal D}X^{10}\;{\cal D}X^{01}\;exp\;i \int dt\; 
L_{eff}(X,\dot{X}) \;,
\eqnum{2.15}
\end{equation}

\noindent
where

\begin{equation}
L_{eff}(X,\dot{X}) = \frac{i}{2}(X^{01} {\dot X^{10}} + X^{10} {\dot X^{01}})
- H(X)\mid_{\Omega_{a} = 0}\;.
\eqnum{2.16}
\end{equation}

As it can be seen, this model initially has two bosonic degrees of freedom and 
the FJ algorithm provides two bosonic constraints, so the bosonic dynamics is 
lost. Therefore, Eq. (2.15) is only dependent on one complex Grassmann
variable, and it is precisely the path-integral representation for the 
partition function of the spinless fermion model.
 
This example shows how our approach produces the correct effective 
Lagrangian for the model.

\section{Family of first-order Lagrangian and constraints in the} ${t-J}$ 
MODEL.   

In the usual approach to study the quantization problem in the $t-J$ model 
through the path-integral representation both, the slave-fermion and the
slave-boson representations are available\cite{3}. In these representations 
the real excitations are forced to be decoupled. We must note that 
even using the slave-particle representations, the constrained Dirac theory is 
needed\cite{13}.

Recently, by using the t-J model to solve a particular problem of
superconductivity, a discrepancy between the results of the
slave-boson and the $X$-operator approaches was found\cite{14}. Such a 
discrepancy is probably an artifice of the slave representation. From our 
point of
view this situation is important and must be taken into account since we 
behave in such a way that the Hubbard $X$-operators are treated as  
indivisible objects.
  
Another alternative way without any decoupling assumption, is to study the 
system from the point of view of the coherent state phase path-integration
\cite{15}.  

In this section following our approach, the Lagrangian dynamics 
generated by the most general graded algebra spl(2,1) is constructed. 
In this case, the four
quantities $(X^{+-}\;,\;X^{-+}\;,\;X^{++}\;,\;X^{--})$ are boson-like operators
and the four quantities $(X^{0+}\;,\;X^{0-}\;,\;X^{+0}\;,\;X^{-0})$ are
fermion-like. As we will see later on, in our approach the remaining 
boson-like operator $X^{00}$ is really a function of the fermion-like 
operators.   

The Lagrangian (2.1) for the t-J model explicitly reads

\begin{eqnarray}
L & = & a_{+ -}(X)\;{\dot{X}}^{+ -} + a_{- +}(X)\;{\dot{X}}^{- +} + 
a_{+ +}(X)\;{\dot{X}}^{+ +} + a_{- -}(X)\;{\dot{X}}^{- -} + 
a_{0 +}(X)\;{\dot{X}}^{0 +} \nonumber \\ 
& + & a_{0 -}(X)\;{\dot{X}}^{0 -} 
+ a_{+ 0}(X)\;{\dot{X}}^{+ 0} + a_{- 0}(X)\;{\dot{X}}^{- 0} - {\bf V^{(0)}}\;.
\eqnum{3.1}
\end{eqnarray} 

By defining $a_{u} = \frac{1}{2}(a^{+ +} - a^{- -})$ and $a_{v} = 
\frac{1}{2}(a^{+ +} + a^{- -})$, and by calling 
$u = X^{++} - X^{--}$ and $v = X^{++} + X^{--}$ the Lagrangian (3.1) can be
written in the more useful form
 
\begin{eqnarray}
L & = & a_{+ -}(X)\;{\dot{X}}^{+ -} + a_{- +}(X)\;{\dot{X}}^{- +} + 
a_{u}(X)\;{\dot{u}} + a_{v}(X)\;{\dot{v}} \nonumber \\ 
& + & a_{0 +}(X)\;{\dot{X}}^{0 +} + a_{0 -}(X)\;{\dot{X}}^{0 -} 
+ a_{+ 0}(X)\;{\dot{X}}^{+ 0} + a_{- 0}(X)\;{\dot{X}}^{- 0} \nonumber \\ 
& - & {\bf V^{(0)}}\;.
\eqnum{3.2}
\end{eqnarray} 

Analogously to that developed above, also in the general case of the graded 
algebra spl(2,1) the symplectic supermatrix $M_{AB}(X)$ can be constructed
straightforward. The starting symplectic supermatrix $M_{AB}(X)$ is 
again singular, and one iteration is necessary to obtain an invertible 
supermatrix.   

Taking into account that there are only four bosonic fields, only two bosonic 
constraints are possible (see discussion in Ref.[1]). Therefore, 
the antisymmetric ordinary bosonic submatrix $A_{bb}$ given in Eq. (2.8) must 
be a $6\times6$ dimensional square matrix.

In the t-J model case, the system of Eqs.(2.10) on the constraints
is given by eight homogeneous differential equations. Such a system has two
bosonic solutions, and the corresponding constraints writes   

\begin{equation}
\Omega_{1} = X^{++} + X^{--} + \rho - 1 = 0\;, 
\eqnum{3.3a}
\end{equation}
 
\begin{equation}
\Omega_{2} = X^{+ -} X^{- +} + \frac{1}{4} u^{2} - (1 - \frac{1}{2}v)^{2}
+ \rho = 0\;, 
\eqnum{3.3b}
\end{equation}

\noindent
where $\rho$ is defined by

\begin{equation}
{\rho} = X^{0+} X^{+0} + X^{0-} X^{-0}\;. 
\eqnum{3.4}
\end{equation}

It is important to notice that in the present case the constraint (3.3a) 
solves the differential 
equations system (2.10), if and only if the following fermionic 
constraints hold  

\begin{equation}
\Xi_{1} = X^{0+} X^{--} - X^{0-}X^{-+} = 0\;, 
\eqnum{3.5a}
\end{equation}

\begin{equation}
\Xi_{2} =  X^{+0} X^{--} - X^{-0}X^{+-} = 0\;, 
\eqnum{3.5b}
\end{equation}

\begin{equation}
\Xi_{3} =  X^{0+} X^{+-} - X^{0-}X^{++} = 0\;, 
\eqnum{3.5c}
\end{equation}

\begin{equation}
\Xi_{4} =  X^{+0} X^{-+} - X^{-0}X^{++} = 0\;. 
\eqnum{3.5d}
\end{equation}

Of course, the four fermionic constraints (3.5), are also solutions of 
the differential equations system.

As it can be seen only two of the fermionic constraints (3.5) must be
considered as independent. In fact, from Eqs. (3.5c), (3.5d) and using 
the Eq.(3.3b), it is easy to show that the Eqs.(3.5a) and (3.5b) can be 
recovered.

Therefore, in the t-J model under consideration, there are 
two bosonic constraints given by Eqs. (3.3) and two
fermionic constraints given by Eqs. (3.5c) and (3.5d). 

In such a condition the symmetric non-singular 
ordinary bosonic submatrix  $D_{ff}$ also results a $6\times6$ dimensional
square matrix. 
Thus, the symplectic supermatrix written in Eq. (2.8) has dimension 
$12\times12$.

As in the spinless fermion case, in the t-J model  
the completeness condition (1.3) is also obtained as one of the bosonic 
constraints (Eq.[3.3a]). So, in Eq. (3.3a) 
$\rho$ must be identified with the hole density (i.e., proportional to the 
number of holes $X^{00}$).

We remember that such a condition has an important physical 
meaning, and it must be imposed to avoid at quantum level the configuration 
with double occupancy at each site. 

Therefore, we must emphasize that by means of our approach the completeness 
condition appears as necessary by consistency.

The next step is to determine the Lagrangian coefficients functions 
$a_{\alpha \beta}(X)$ appearing in the equation (3.2).
This is straightforward
by using $M_{AB}(M^{BC})^{-1} = \delta^{C}_{A}$, and solving the system of
partial differential equations on such coefficients. 
The algebraic manipulations are rather 
similar to that given in the pure bosonic case (see Ref.[1]) but more 
complicated. So, we only write here the final results.

After some algebra, it can be shown that a family of solutions of the partial 
differential equations system can be written as follows

\noindent
a) For the bosonic coefficients:

\begin{equation}
a_{+ -} = i\; F(u,v,\rho)\; X^{- +}\; ,
\eqnum{3.6a}
\end{equation}

\begin{equation}
a_{- +} =  a^{*}_{+ -} = - i\; F(u,v,\rho)\; X^{+ -}\; ,
\eqnum{3.6b}
\end{equation}

\noindent
and $a_{u}$ , $a_{v}$ are respectively arbitrary functions of
the $u$ and $v$ variables. For simplicity these two last coefficients 
can be also taken equal to zero.

\noindent
b) For the fermionic coefficients:

\begin{equation}
a_{- 0} = \frac{i}{2} X^{0 -}\; ,
\eqnum{3.6c}
\end{equation}

\begin{equation}
a_{0 -} = \frac{i}{2} X^{- 0}\; ,
\eqnum{3.6d}
\end{equation}

\begin{equation}
a_{+ 0} = \frac{i}{2} X^{0 +}\; ,
\eqnum{3.6e}
\end{equation}

\begin{equation}
a_{0 +} = \frac{i}{2} X^{+ 0}\; .
\eqnum{3.6f}
\end{equation}

The real function $F(u,v,\rho)$ appearing in Eqs. (3.6a,b) verifies the 
following system of partial differential equations  

\begin{equation}
4 X^{+ -} X^{- +} \frac{\partial F}{\partial u} - 2 F u = 1 + \rho\; ,
\eqnum{3.7a}
\end{equation}

\begin{equation}
\left[2 X^{+ -} X^{- +}\left(\frac{\partial F}{\partial X^{- -}} - 
\frac{\partial F}{\partial \rho}\right) + 2 X^{+ +} F + X^{+ +}\right]
X^{0 +} = 0\; ,
\eqnum{3.7b}
\end{equation}

\begin{equation}
\left[2 X^{+ -} X^{- +}\left(\frac{\partial F}{\partial X^{+ +}} - 
\frac{\partial F}{\partial \rho}\right) + 2 X^{- -} F - X^{- -}\right]
X^{0 +} = 0\; ,
\eqnum{3.7c}
\end{equation}

\begin{equation}
2 F \rho +  X^{+ -} X^{- +}\rho \left[\frac{1}{X^{+ +}}\left(\frac{\partial F}
{\partial X^{- -}} - \frac{\partial F}{\partial \rho}\right) + 
\frac{1}{X^{- -}}\left(\frac{\partial F}
{\partial X^{+ +}} - \frac{\partial F}{\partial \rho}\right)\right] = 0\; .
\eqnum{3.7d}
\end{equation}

A family of solutions of the system (3.7) which has physical interest is given 
by

\begin{equation}
F(u,v,\rho) = \frac{(1 + \rho)u + \alpha}{(2 - v)^{2} - 4\rho - u^{2}}\;,
\eqnum{3.8} 
\end{equation}

\noindent 
being $\alpha$ an arbitrary and non trivial integration constant. 

In the bosonic limit when $\rho = 0$ and $v = 1$ it results 

\begin{equation}
F(u) = \frac{u + \alpha}{1 - u^{2}}\;.
\eqnum{3.9} 
\end{equation}

\noindent
recovering the solution for the pure bosonic case (see Refs.[1,16]). 

For simplicity, we choose a particular family of solutions by taking 
$a_{u} = a_{v} = 0$ and $\alpha = -1$, so the Lagrangian (3.2) without 
accounting the constraints can be written 

\begin{eqnarray}
L(X,\dot{X}) & = & i\sum_{i}\frac{(1 + \rho_{i})u_{i} - 1}{(2 - v_{i})^{2} 
- 4\rho_{i} - u^{2}_{i}}
\left(X^{- +}_{i} {\dot X}^{+ -}_{i} - X^{+ -}_{i} {\dot X}^{- +}_{i}\right)
\nonumber \\ 
& + & \frac{i}{2} \sum_{i, \sigma} \left(X_{i}^{\sigma 0}{\dot X}_{i}^{0 
\sigma} + X_{i}^{0 \sigma}{\dot X}_{i}^{\sigma 0}\right) - H_{t-J}(X)\;,
\eqnum{3.10}
\end{eqnarray}

\noindent
where $\sigma$ takes the values $+$ and $-$, and the site index $i$ was added.

In Eq. (3.10) the well known Hamiltonian $H_{t-J}$ for the t-J model
in terms of the Hubbard operators is given by
  
\begin{equation}
H_{t-J} =  \sum_{i,j,\sigma} \; t_{ij}\; X^{\sigma 0}_{i}
X^{0 \sigma }_{j} + \frac{1}{4} \sum_{i,j, \sigma, \bar{\sigma}}
J_{ij}\; X^{\sigma {\bar{\sigma}}}_{i}
X^{\bar{\sigma} \sigma}_{j} - \frac{1}{4} \sum_{i,j, \sigma, \sigma'}
J_{ij}\; X^{\sigma \sigma'}_{i}
X^{\bar{\sigma}{\bar{\sigma'}}}_{j}\;. 
\eqnum{3.11}
\end{equation}

At this stage we are ready to carry out the quantization of the model 
by using functional techniques.

\section{Path integral representation in terms of a real vector field.}

Now, we go on writing the correlation generating functional for the t-J 
model by using again the path-integral Faddeev-Senjanovich approach. 
In terms of the Hubbard $X$-operators the partition function can be formally 
written as follows 

\begin{equation}
Z = \int {\cal D}X_{i}\; \delta(\Omega_{i1})\;
\delta(\Omega_{i2})\; \delta(\Xi_{i3})\;\delta(\Xi_{i4})\; 
sdet M_{AB}\;  exp\;i \int dt\; L(X,\dot{X}) \;,
\eqnum{4.1}
\end{equation}

\noindent
where $L(X,\dot{X})$ is given in Eq. (3.10) and the constraints 
$\Omega_{1}$, $\Omega_{2}$, $\Xi_{3}$ and $\Xi_{4}$ are given in 
Eqs. (3.3a,b), (3.5c) and (3.5d) respectively. 

We note that the function $sdet M_{AB}$ appearing in equation (4.1),  
is the superdeterminant of the symplectic supermatrix (2.8). Really, in the
path-integral formalism of Faddeev-Senjanovich such superdeterminant 
correspond to the supermatrix constructed from the
second-class constraints provided by the Dirac formalism. 
As it can be shown both, the symplectic supermatrix and the supermatrix
constructed from the second-class constraints 
are equals\cite{6}. 
 
Such superdeterminant is computed, and after some algebra we get

\begin{equation}
sdet M_{AB} = det A \left[det(D - CA^{-1}B)\right]^{-1} = 
- \frac{4(1 + \rho)^{2}}{(1 - \rho + u)^{2}} \; ,
\eqnum{4.2}
\end{equation}

\noindent
where $A$, $B$, $C$ and $D$ are the submatrices defined in (2.8).

At this stage, it is useful to relate the 
boson-like $X$-Hubbard operators with the real
components $S_{i}$ ($i = 1,2,3$) of a vector field ${\bf S}$,
by means of a transformation. 
By using explicitly the constraint (3.3a), the independent bosonic
degrees of freedom  are reduced to three, and we can write

\begin{equation}
X^{+ +} = \frac{1}{2 s} (1 - \rho)(s + S_{3}) \;,
\eqnum{4.3a} 
\end{equation}

\begin{equation}
X^{- -} = \frac{1}{2 s} (1 - \rho)(s - S_{3}) \;,
\eqnum{4.3b} 
\end{equation}

\begin{equation}
X^{+ -} = \frac{1}{2 s} (1 - \rho)(S_{1} + iS_{2}) \;,
\eqnum{4.3c} 
\end{equation}

\begin{equation}
X^{- +} = \frac{1}{2 s} (1 - \rho)(S_{1} - iS_{2}) \;,
\eqnum{4.3d} 
\end{equation}

\noindent
where $s$ is a constant.

Note that only when $\rho = 0$ (pure bosonic case), the real vector field 
${\bf S}$ can be identified with the spin.
 
Moreover, the fermion-like $X$-Hubbard operators can be related with 
Grassmann variables i.e., suitable component spinors

\begin{equation}
X^{- 0} =  \Psi_{+}  \hspace{2cm} X^{0 -} =  \Psi_{+}^{*}\;,
\eqnum{4.4a} 
\end{equation}

\begin{equation}
X^{+ 0} = \Psi_{-}  \hspace{2cm} X^{0 +} = \Psi_{-}^{*}\;,
\eqnum{4.4b} 
\end{equation}
 
\noindent
where now  $\rho = \Psi_{+}^{*} \Psi_{+} + \Psi_{-}^{*} \Psi_{-}$ and
accounting the fermionic constraints (3.5) it results 
$(1 - \rho)(1 + \rho) = 1$.

The remaining bosonic constraint (3.3b) as function of the real vector field 
variable ${\bf S}$ writes

\begin{equation}
\Omega_{2} = S_{1}^{2} + S_{2}^{2} + S_{3}^{2} - s^{2} = 0\;.
\eqnum{4.5}
\end{equation}

Analogously, the two fermionic constraints (3.5c) and (3.5d) can be written

\begin{equation}
\Xi_{3} = \Psi_{-}^{*}(S_{1} + iS_{2}) - \Psi_{+}^{*}(s + S_{3}) = 0\;,
\eqnum{4.6a}
\end{equation}

\begin{equation}
\Xi_{4} = \Psi_{-}(S_{1} - iS_{2}) - \Psi_{+}(s + S_{3}) = 0\;.
\eqnum{4.6b}
\end{equation}
 
Consequently, by neglecting a total time derivative 
the Lagrangian (3.10) in terms of these new fields reads

\begin{equation}
L = - \frac{1}{2s} \sum_{i} \frac{S_{i2} \dot{S_{i1}} - S_{i1} 
\dot{S_{i2}}}{s + S_{i3}}
+ i \sum_{i,\sigma}\Psi_{i,\sigma}{\dot{\Psi}^{*}_{i,\sigma}} - 
H_{t-J}\;.
\eqnum{4.7}
\end{equation}

The Hamiltonian $H_{t-J}$ given in (3.11) written in term of the real vector 
variable ${\bf S}$ and the component spinors (4.4), takes the form

\begin{equation}
H_{t-J} =  \sum_{i,j,\sigma} t_{ij} \Psi_{i \sigma}\Psi_{j \sigma}^{*}
+\frac{1}{8 s^{2}}\sum_{i,j} J_{ij}(1 - \rho_{i})(1 - \rho_{j}) \left[
S_{i1}S_{j1} + S_{i2}S_{j2} + S_{i3}S_{j3} - s^{2}\right]\;.  
\eqnum{4.8}
\end{equation}

Finally, we note that the fermionic constraints (3.5) can be written in 
matrix notation as follows

\begin{equation}
\Xi = \left(s{\bf{I}} + {\bf S}.{\mbox{\boldmath{$\sigma$}}}\right)
{\bf{\Psi}} = 0\;,
\eqnum{4.9a}
\end{equation}

\begin{equation}
\Xi^{*} =  {\bf{\Psi}}^{*}\left(s{\bf{I}} + {\bf S}.
{\mbox{\boldmath{$\sigma$}}}\right) = 0
\;,
\eqnum{4.9b}
\end{equation}

\noindent
where
${\mbox{\boldmath{$\sigma$}}}$ are the Pauli matrices, and the 
two-component spinor ${\bf{\Psi}} = (- {\Psi}_{+}\;,\;{\Psi}_{-})$ is 
defined. The fermionic constraints (4.9)
are precisely those used in Ref. [15] in the framework of the coherent
state representation.

The Eq. (4.1) can be written in an alternative way by using the integral 
representation for the delta functions on the constraints $\Phi$ 

\begin{eqnarray*}
\delta({\Phi}) = \int {\cal D}{\chi} \; exp \;(i \;\int\; dt\; 
{\chi}\;{\Phi}) \; ,
\end{eqnarray*}

\noindent
where the quantities
${\chi}$ are suitable bosonic or fermionic Lagrange multipliers.
 
Consequently, the correlation generating functional (4.1) takes the form

\begin{equation}
Z  =  \int {\cal D}X_{i}\; {\cal D}{\lambda_{i}^{1}}\; 
{\cal D}{\lambda_{i}^{2}}\; 
{\cal D}{\xi_{i}}\; {\cal D}{\xi^{*}_{i}}\; 
\; sdet M_{AB}\; exp\;i \int dt\; L_{eff}(X,\dot{X}) \;.
\eqnum{4.10}
\end{equation}

The effective Lagrangian $L_{eff}(X,\dot{X})$ appearing in Eq. (4.10)
is defined by
 
\begin{eqnarray}
L_{eff}(X,\dot{X}) & =  & L(X,\dot{X}) + \sum_{i} \lambda_{i}^{1}\left(
X^{++}_{i} + X^{--}_{i} + \rho_{i} - 1 \right) \nonumber \\
& + & \sum_{i} \lambda_{i}^{2}\left[X^{+ -}_{i} X^{- +}_{i} 
+ \frac{1}{4} u^{2}_{i} - (1 - \frac{1}{2}v_{i})^{2} 
+ \rho_{i}\right] \nonumber \\
& + & \left( X^{0+}_{i} X^{+-}_{i} - X^{0-}_{i} X^{++}_{i}\right)
{\xi_{i}}  
+ {\xi}^{*}_{i} \left(X^{+0}_{i} X^{-+}_{i} - X^{-0}_{i} X^{++}_{i}\right)
\;.   
\eqnum{4.11}
\end{eqnarray}

\noindent
where $L(X,\dot{X})$ was given in Eq. (3.10).

In Eq. (4.11) the parameters $\lambda^{a}$ ($a = 1,2$) and ${\xi}$ 
are respectively bosonic and fermionic Lagrange multipliers. 

The functional $sdet M_{AB}$ written in terms of the real vector field 
${\bf S}$ and the two-component spinor ${\bf{\Psi}}$ results  

\begin{equation}
sdet M_{AB} =  - \frac{4s^{2}(1 + \rho)^{2}}{(1 - \rho)^{2}(s + S_{3})^{2}} 
\eqnum{4.12}
\end{equation}

The exponentiation of the superdeterminant    
is realized as usual by introducing the Faddeev-Popov ghosts in the 
effective Lagrangian. Therefore, we assume that the functional $sdet M_{AB}$ 
is written as follows

\begin{equation}
sdet M = \int\; {\cal D}C\; exp\; i C_{1}^{A} M_{AB} C_{2}^{B}\;,
\eqnum{4.13}
\end{equation}

\noindent
where ${\cal D}C = \prod_{A} {\cal D}C_{2}^{A}\;{\cal D}C_{1}^{A}$ and 
$C^{A}_{a}$ ($a = 1,2$) denote commuting as well as anticommuting 
ghosts.

It is useful to analyze the partition function (4.10) when it is written 
in terms of the real vector field variable ${\bf S}$. 

Making use of the transformation (4.3) and (4.4), Eq. (4.10) takes the form 

\begin{equation}
Z  =  \int  {\cal D} S_{i1} \; {\cal D} S_{i2}\; {\cal D} S_{i3}\; 
{\cal D}{\Psi}_{i \sigma}\; 
{\cal D}{\Psi}^{*}_{i \sigma}\; 
{\cal D}{\lambda_{i}^{2}}\; {\cal D}{\xi_{i}}\; {\cal D}{\xi_{i}^{*}}\;
sdet M_{AB}\;
\frac{\partial X}{\partial S}\;
exp\;(i \int\; dt \; L_{eff})\;.
\eqnum{4.14}
\end{equation}

\noindent 
where the quantity $\frac{\partial X}{\partial S}$ is the super Jacobian of 
the transformation (4.3 - 4.4), also a field dependent functional whose value 
is 

\begin{equation} 
\frac{\partial X}{\partial S} = - i \frac{(1 - \rho)^{3}}{2 s^{3}}\;.
\eqnum{4.15}
\end{equation}

The effective Lagrangian $L_{eff}$ given in Eq. (4.14), in terms of the new
variables reads  
 
\begin{eqnarray} 
L_{eff} & = & \frac{1}{2s} \sum_{i} \frac{S_{i1}{\dot S}_{i2} 
- S_{i2}{\dot S}_{i1}}{s + S_{i3}} + i \sum_{i, \sigma} 
{\Psi_{i \sigma}}{\dot{\Psi}^{*}}_{i \sigma} - H_{t-J} \nonumber \\
& + & \sum_{i}\left[\lambda_{i}^{2}(S_{i1}^{2} + S_{i2}^{2} + S_{i3}^{2} 
- s^{2})
+ \xi^{*}_{i} \left(\Psi_{-}(S_{i1} - iS_{i2}) - \Psi_{+}(s + S_{i3})
\right) \right. \nonumber \\ 
& + & \left. \left(\Psi_{-}^{*}(S_{i1} + iS_{i2}) - \Psi_{+}^{*}(s + S_{i3})
\right) \xi_{i} \right] \;.
\eqnum{4.16}
\end{eqnarray}

As a last comment we must say that the treatment of the path-integral (4.14)
is cumbersome. 

By one hand, the effective Lagrangian (4.16) depends on the
hole density $\rho$. When there is a small number of holes it can be assumed
that the hole density $\rho = constant$. In this
situation the super Jacobian of the transformation (4.3 - 4.4) is constant and
it contributes only to the normalization factor of the path-integral (4.14). 

On the other hand, the non-polynomial structure of the effective Lagrangian 
(4.16) is due to the contribution of both, the bosonic kinetic part and the 
terms coming
from the functional $sdet M$. This problem is also present in the pure bosonic
case (see for instance Refs. [1,16]), and is solved in the framework of the
perturbative formalism. The non-polynomial character of $L_{eff}$ is due
to the presence of the component $S_{3}$ of the real vector field
${\bf S}$ in the denominator. So, this problem can be treated by considering 
the effective Lagrangian fluctuating around the antiferromagnetic background. 

In these conditions, and at least when only first-order terms of the 
perturbative development are retained the $sdet M$
is constant, and so it is possible to obtain results without introducing 
ghosts.

In a forthcoming paper under preparation, by following these prescriptions
the non-polynomial effective Lagrangian is studied. On the basis of our 
path-integral formulation and by applying the perturbative formalism 
the Feynmann rules and the diagrammatics of the t-J model will be given.    

\section{Conclusions}

In this paper a discussion about the construction of a family of
first-order Lagrangian describing the dynamics of the t-J model is presented.
In this approach any decoupling is used, but 
the field variables are directly the Hubbard $X$-operators 
satisfiying the graded algebra spl(2,1). 
Using the supersymmetric version of the Faddeev-Jackiw symplectic formalism, 
we have shown that it is possible to find a family of first-order Lagrangian
able to reproduce at classical level the generalized FJ brackets or graded 
Dirac brackets of the t-J model.
When the transition to the quantum theory is realized as usual in a
canonical quantum formalism, the graded quantum Dirac brackets are precisely 
the graded commutators of the Hubbard $X$-operators algebra. 
Moreover, in both cases the spinless
fermion model as well as the t-J model, the unique 
possible set of constraints is naturally provided by the
symplectic FJ method.

From our approach applied to the simple case of the graded algebra spl(1,1) 
(i.e, the spinless fermion model), we have shown that the partition function 
can be written in terms of the same effective Lagrangian obtained by means of 
other methods. 

Also the t-J model is treated in the framework of the path-integral 
representation by using the Hubbard $X$-operators  
as field variables. In this context the correlation generating funtional
is constructed. 

Later on, by making  a transformation from the boson-like Hubbard 
$X$-operators 
to a real boson vector field ${\bf S}$, we have rewritten the effective 
Lagrangian appearing in the partition function. The real vector field 
${\bf S}$ has the particularity that in the bosonic limit i.e., when 
the fermion-like operators are withdrawed, it is not other than the spin 
vector field. As it is also shown, in the bosonic limit the remaining bosonic 
part of our non-polynomial Lagrangian 
is equal to that given in Refs. [1,16], checking in this way 
the expression of the partition function for the pure bosonic case.
 
In summary, we can conclude that starting from a total independent scenario, 
the path-integral representation we found is equivalent to that 
obtained by means of the coherent states method. 

We think that it would be interesting to compare the present 
formulation with those that could be obtained by using the coherent state 
quantization of constrained systems formalism recently developed in Ref. [17].

Finally, it must be noted (see Ref. [1]) that among the solutions we have 
found, 
also the solution given in Ref. [15] for the bosonic kinetic part is obtained.
This is done by defining an appropriate vector ${\bf a}$ which verifies
the equation $({\nabla\times{\bf a}}){\bf S} = 1 + \rho$.
However we believe that the bosonic solution we choose to write the 
Lagrangian (4.7) is more convenient for our future purposes.

\noindent
{\bf Acknowledgments}

\noindent
The authors want to acknowledge to D. F\"{o}sters for his suggestion about
checking our method on the spinless fermion model. A. G. thanks 
to A. Dobry and R. Zeyher for useful discussions.

\end{document}